\documentclass{article}
\usepackage{PRIMEarxiv}

\usepackage[inline]{enumitem}
\usepackage{pbox}
\usepackage{balance}
\usepackage{todonotes}
\newlist{inlinelist}{enumerate*}{1}
\setlist*[inlinelist,1]{label=\roman*),itemjoin={{, }},itemjoin*={{, and }}}
\usepackage{url}
\usepackage[normalem]{ulem}
\usepackage{multirow}
\usepackage{multicol}
\usepackage{booktabs}
\usepackage{amsfonts}
\usepackage{amsmath}
\usepackage{amssymb}
\usepackage{amsthm}
\usepackage{adjustbox}
 \usepackage[normalem]{ulem}
 \useunder{\uline}{\ul}{}

\title{Naver Labs Europe (SPLADE) @ \\ TREC NeuCLIR 2022}

\author{
     Carlos Lassance, Stephane Clinchant \\
  Naver Labs Europe \\
  France\\
  \texttt{carlos.lassance@naverlabs.com, stephane.clinchant@naverlabs.com}
}

\begin{document}

\maketitle

\begin{abstract}
This paper describes our participation in the 2022 TREC NeuCLIR challenge. We submitted runs to two out of the three languages (Farsi and Russian), with a focus on first-stage rankers and comparing mono-lingual strategies to Adhoc ones. For monolingual runs, we start from pretraining models on the target language using MLM+FLOPS and then finetuning using the MSMARCO translated to the language either with ColBERT or SPLADE as the retrieval model. While for the Adhoc task we test both query translation (to the target language) and back-translation of the documents (to english). Initial result analysis shows that the monolingual strategy is strong, but that for the moment Adhoc achieved the best results, with back-translating documents being better than translating queries.

\end{abstract}

\section{Introduction}
In this paper, we detail our TREC 2022 NeuCLIR track submission, based on the latest improvements of the SPLADE model~\cite{efficiency,pp}, with the main factor being the MLM+FLOPS pretraining (for monolingual runs) and distillation (for the Adhoc runs). In total, we submitted 7 runs per language, being 3 baselines based solely on SPLADE (monolingual, Adhoc via query translation and Adhoc via document back-translation\footnote{Throughout this notebook we use back-translation as the translation from the target language to English}) and 4 main runs (divided into monolingual/Adhoc and reranked/ensemble of the first stage rankers). 

Compared to our TREC DL 2021 notebook, we decided to make this notebook more streamlined compared to last year (and more drafty). For a more thorough introduction of the models used here, we invite the reader to check the following articles: SPLADE training~\cite{pp}, MLM+FLOPS pretraining~\cite{efficiency}, ColBERT\cite{colbert}, Rocchio~\cite{rocchio}, Training style we used for our rerankers~\cite{gao2021rethink} and T5 based reranking~\cite{nogueira2020document}.

\section{Methodology}

In the following, we introduce the models we consider for both candidate generation as well as re-ranking. We also describe our training procedure and detail the submitted runs. SPLADE and pretrained language models are made available at \url{https://huggingface.co/naver/\{name\}} with the models being named: neuclir22-\{pretrained, splade\}-\{fa,ru,zh\}. Note that even if we did not participate in zh, we trained models a posteriori and made them available.

\subsection{First Stage}

For the first stage this year we separate models into two avenues: trained on the target language and on English. For the latter we always use the Splade++ CoCondenser Self Distil\footnote{Available at \url{https://huggingface.co/naver/splade-cocondenser-selfdistil}} and for the former:
\begin{enumerate}
    \item We pre-train 6L DistilBERT base models from scratch using MLM+FLOPS~\cite{pretraining} on the target language using a combination of the target documents, translated MSMARCO~\cite{bonifacio2021mmarco} and Mr.Tydi~\cite{zhang2021mr} depending on their availability for the language.
    \item We then finetune those models with ColBERT~\cite{colbert} and SPLADE~\cite{pp} style training on the translated MSMARCO~\cite{bonifacio2021mmarco}.
    \begin{enumerate}
        \item For SPLADE we use the negatives from \url{https://huggingface.co/datasets/sentence-transformers/msmarco-hard-negatives} drawn from a multitude of models on the English version of the corpus. Note that we tried using distillation with the English scores (which we already had tested on other languages with positive results), but the better results came from non-distillation training. 
        \item While for ColBERT we use the traditional English BM25 negatives. In order to maximize ColBERT performance we remove the last layer (which reduces the dimensionality from 768 to 128) and inference is done using brute-force retrieval instead of an approximate ANN (following ~\cite{lassance2021colbert}).
    \end{enumerate}
    \item For completeness, we also include BM25 in our first stage ensembling. However, in hindsight, we should have used the run provided by the organizers as the last piece of the ensemble.
\end{enumerate}

\subsection{Second Stage}

As per last year's competition we use a mix of different PLMs as rerankers for which training is inspired by~\footnote{code available at \url{https://github.com/luyug/Reranker}}~\cite{gao2021rethink} and using negatives from SPLADE. For the monolingual runs we used InfoXLM and XLM-Roberta-Large as the PLMs, while for the Adhoc runs we consider only rerankers trained on english, using Electra-Large, Deberta-v3-Large and Deberta-v2-xxLarge\footnote{Which are similar to the ones used on TREC-DL22 and made available on huggingface} (and thus apply them using the back-translated documents). We also add the pretrained MonoT5-3B ~\cite{nogueira2020document}.

\subsection{Ensembling}

We also have applied ensembling in order to improve our results. This year we used ranx~\cite{bassani2022ranx} to generate all our ensembles, using average normalized score over the ensembles, unless explicitly noted. The normalized score uses the min and max values of the query so that, for each model, the best score is 1 and the lowest one 0.

\subsection{Runs submitted to TREC}

We submitted a total of 14 runs, 7 for Farsi and 7 for Russian. Of the 7 runs, we submitted 3 baselines (human query translation, machine query translation, machine document translation) and 4 main runs (combination of monolingual/Adhoc with first-stage only/reranked). Namely:

\begin{enumerate}
    \item splade\_\{language\}\_ht: Baseline monolingual first-stage ranking with SPLADE and human translated queries
    \item splade\_\{language\}\_mt: Baseline Ad-hoc first-stage ranking with SPLADE and machine-translated queries
    \item splade\_\{language\}\_dt: Baseline Ad-hoc first-stage ranking with SPLADE and machine-translated documents
    \item NLE\_\{language\}\_mono: Monolingual ensemble of first-stage rankers on human translated queries
    \item NLE\_\{language\}\_mono\_rr: Monolingual ensemble of first-stage and rerankers on human translated queries
    \item NLE\_\{language\}\_adhoc: Ad-hoc ensemble of first-stage rankers on machine-translated queries and/or machine-back-translated documents
    \item NLE\_\{language\}\_adhoc\_rr: Ad-hoc ensemble of first-stage and rerankers on machine-translated queries and/or machine-back-translated documents\end{enumerate}

\section{Analysis on Filtered HC4}
In order to generate our final runs, we needed to set a development set. During the challenge period, we tested many different development sets, but in the end, we focused on what we called \emph{HC4-filtered dev}. The \emph{HC4-filtered dev} is the HC4~\cite{Lawrie2022HC4} dev set, filtered to include only the documents present in the competition's test set (NeuCLIR1)\footnote{This include both in the collection and in the qrels}. Other possibilities we rejected were:\begin{itemize}
    \item MSMARCO translated dev set: While MSMARCO is a good avenue for training, we were afraid that the translated dev set could be biased to the translation and thus make the results less viable (we also would not have a human translated development set).
    \item Mr.TyDi dev/test set: Discarded because it is not available in Farsi.
    \item Full HC4 dev set: Discarded due to its size (larger than the filtered one) and the fact that it did not focus on the same documents as the test set.
\end{itemize}

We shared our results on the HC4-filtered dev with other participants and it seemed like a solid baseline, we thus shared some run files so that other participants could use our SPLADE as their first-stage rankers. Results are made available in Table~\ref{tab:monolingual} (monolingual) and Table~\ref{tab:adhoc} (Adhoc), from which we take the following initial conclusions:

\begin{itemize}
    \item In almost all considered tasks we had a hierarchy of SPLADE $\ge$ ColBERT $\ge$ BM25, but their ensemble seemed more stable than individual runs (even if not always better).
    \item However the precision boost for first-stage rankers over BM25 vary depending on language, while it is almost double in Farsi, in Russian it is of only a few points. However, in terms of Recall, there's a larger gap.
    \item Also, we noticed that SPLADE+Rocchio almost always improved over SPLADE by itself, something we also noticed on the TREC DL challenge. 
    \item Differently from English, monolingual runs did not always improve after reranking. However, the ensembling of rerankers and first-stage rankers was always better than the first-stage ranking by itself. One thing that we missed is that we could have used our first-stage rankers to initialize the rerankers, which could have improved the results.
    \item The best AdHoc ensemble varied by language, with Russian using all available models and Farsi focusing only on the T5 reranked runs. In hindsight, this might not have been the best decision.
    \item Continuing on possibly bad decisions, the first stage AdHoc ensembles did not agree if we should include or not the document back translated ones. 
    \item The results seem to indicate that monolingual vs AdHoc seems to be language and stage dependent. For first stage, Farsi obtained the best results on monolingual, while Russian was on Adhoc (using back-translated documents). For reranking results were always better using back-translated documents, which is expected because the PLMs are not as good (and hyperparameters have not been tuned as well).
    \item It would have been nice to compare with SPLADE-X~\cite{nair2022learning} a CLIR model that came out after the competition (and one of its baselines PSQ). However, the authors only compared on CLEF datasets and as far as we are aware the models are not available.
\end{itemize}

\begin{table}[ht]
\centering
\label{tab:monolingual}
\caption{MonoLingual results on HC4-filtered dev. Lines containing submissions are bolded. nDCG@20 multiplied by 100 for ease of presentation.}
\adjustbox{max width=\textwidth}{%
\begin{tabular}{c|cc|ccc|ccc|c}
\toprule 
\multirow{2}{*}{\#} &
  \multirow{2}{*}{Model} &
  \multirow{2}{*}{Rerank or Ensemble} &
  \multicolumn{3}{c|}{Farsi} &
  \multicolumn{3}{c|}{Russian} &
  \multirow{2}{*}{Name} \\ 
  &                      &             & nDCG@20 & mAP@1k & Recall@1k & nDCG@20 & mAP@1k & Recall@1k &  \\ \midrule
\multicolumn{10}{c}{First stage Rankers}                                                                \\ \midrule
a & BM25                 &             & 19.7    & 13.4   & 79.9\%    & 19.1    & 13.5   & 67.4\%    &  \\
b & SPLADE               &             & 29.8    & 20.4   & 89.5\%    & 21.9    & 16.5   & 75.2\%    &  \\
\textbf{c} &
  \textbf{SPLADE + Rocchio} &
  \textbf{} &
  \textbf{31.6} &
  \textbf{22.1} &
  \textbf{91.4\%} &
  \textbf{22.5} &
  \textbf{16.5} &
  \textbf{70.4\%} &
  \textbf{splade\_\{\}\_ht} \\
d & ColBERT              &             & 30.3    & 21.2   & 89.4\%    & 21.2    & 14.6   & 81.3\%    &  \\ \midrule
\multicolumn{10}{c}{First Stage Ensembles}                                                              \\ \midrule
e & First stage Ensemble & a+c+d       & 31.5    & 21.2   & 91.9\%    & 25      & 17.8   & 78.4\%    &  \\
f & First stage Ensemble & a+b+d       & 30.7    & 20.8   & 91.4\%    & 25      & 18.6   & 78.0\%    &  \\
\textbf{g} &
  \textbf{First stage Ensemble} &
  \textbf{a+b+c+d} &
  \textbf{31.9} &
  \textbf{22.6} &
  \textbf{92.2\%} &
  \textbf{23.8} &
  \textbf{17.4} &
  \textbf{80.7\%} &
  \textbf{NLE\_\{\}\_mono} \\ \midrule
\multicolumn{10}{c}{Rerankers}                                                                          \\ \midrule
h & InfoXLM              & c           & 30.3    & 20.9   & 91.4\%    & 20.1    & 13.1   & 70.4\%    &  \\
i & InfoXLM              & e           & 30.5    & 21.2   & 91.9\%    & 20.7    & 13.6   & 78.4\%    &  \\
j & InfoXLM              & g           & 30.5    & 21.1   & 92.2\%    & 20.6    & 13.5   & 80.7\%    &  \\
k & XLM-Roberta-Large    & c           & 31      & 22.2   & 91.4\%    & 19.9    & 12.7   & 70.4\%    &  \\
l & XLM-Roberta-Large    & e           & 31.4    & 22.1   & 91.9\%    & 20.8    & 13.2   & 78.4\%    &  \\
m & XLM-Roberta-Large    & g           & 31      & 22.2   & 92.2\%    & 20.6    & 13     & 80.7\%    &  \\ \midrule
\multicolumn{10}{c}{Final Ensembles}                                                                    \\ \midrule
n & Final Ensemble       & a+b+c+d+j+m & 34      & 25.5   & 92.2\%    & 25.1    & 17.6   & 80.7\%    &  \\
o & Final Ensemble       & c+d+i+l     & 35.6    & 26.1   & 91.9\%    & 25.8    & 17.8   & 78.4\%    &  \\
\textbf{p} &
  \textbf{Final Ensemble} &
  \textbf{c+d+j+m} &
  \textbf{35.6} &
  \textbf{26.1} &
  \textbf{92.2\%} &
  \textbf{25.8} &
  \textbf{17.8} &
  \textbf{80.7\%} &
  \textbf{NLE\_\{\}\_mono\_rr} \\
q & Final Ensemble       & j+m         & 32.1    & 22.6   & 92.2\%    & 23.1    & 15.2   & 80.7\%    &  \\
r & Final Ensemble       & g+j+m       & 35.4    & 25.7   & 92.2\%    & 25.1    & 17.6   & 80.7\%    & \\ \bottomrule
\end{tabular}
}
\end{table}
\begin{table}[ht]
\centering
\label{tab:adhoc}
\caption{Adhoc results on HC4-filtered dev. Lines containing submissions are bolded. QT: Machine translated queries, DT: Machine translated documents. nDCG@20 multiplied by 100 for ease of presentation.}
\adjustbox{max width=\textwidth}{%

\begin{tabular}{c|cc|ccc|ccc|c}
\toprule
\multirow{2}{*}{\#} &
  \multirow{2}{*}{Model} &
  \multirow{2}{*}{Rerank or Ensemble} &
  \multicolumn{3}{c|}{Farsi} &
  \multicolumn{3}{c|}{Russian} &
  \multirow{2}{*}{Name} \\ 
  &                                    &               & nDCG@20 & mAP@1k & Recall@1k & nDCG@20 & mAP@1k & Recall@1k &                    \\ \midrule
\multicolumn{10}{c}{First Stage Rankers}                                                                                                  \\ \midrule
a & BM25 MT    &               & 17,2    & 11,6   & 81,50\%   & 20      & 14,8   & 68,90\%   &                    \\ 
\textbf{b} &
  \textbf{SPLADE++ MT} &
  \textbf{} &
  \textbf{29,4} &
  \textbf{20,5} &
  \textbf{91,20\%} &
  \textbf{21,6} &
  \textbf{16,8} &
  \textbf{69,90\%} &
  \textbf{splade\_\{\}\_mt} \\
\textbf{c} &
  \textbf{SPLADE++ DT} &
  \textbf{} &
  \textbf{27,4} &
  \textbf{18,3} &
  \textbf{91,50\%} &
  \textbf{23,6} &
  \textbf{16} &
  \textbf{74,20\%} &
  \textbf{splade\_\{\}\_dt} \\
d & ColBERT MT &               & 28,1    & 19,3   & 89,10\%   & 22,4    & 16,3   & 77,30\%   &                    \\ \midrule
\multicolumn{10}{c}{First Stage Ensembles}                                                                                                \\ \midrule
e & Ensemble                           & a+b+c         & 29,3    & 20,4   & 92,00\%   & 26      & 20,4   & 77,00\%   & NLE\_\{\}\_adhoc   \\
f & Ensemble                           & a+b+c+d       & 31,1    & 21,2   & 92,80\%   & 25,8    & 19,7   & 76,70\%   &                    \\ \midrule
\multicolumn{10}{c}{Rerankers over english translated docs}                                                                               \\ \midrule
g & Electra-large                      & f             & 25,2    & 18,3   & 92,80\%   & 18,3    & 11,8   & 76,70\%   &                    \\
h & Deberta-v3-Large                   & f             & 28,3    & 18,9   & 92,80\%   & 22,6    & 13,3   & 76,70\%   &                    \\
i & Deberta-v2-xxLarge                 & f             & 28,4    & 19,8   & 92,80\%   & 25      & 17,4   & 76,70\%   &                    \\
j & T5-3b                              & c             & 36,8    & 27,3   & 91,20\%   & 25      & 19,4   & 69,90\%   &                    \\
k & T5-3b                              & f             & 36,3    & 27,5   & 92,80\%   & 26,1    & 20,6   & 76,70\%   &                    \\ \midrule
\multicolumn{10}{c}{Final ensembles}                                                                                                      \\ \midrule
l & Ensemble                           & a+b+c+d+g+h+i & 33,4    & 23,9   & 92,80\%   & 28,4    & 21,1   & 76,70\%   &                    \\
m & Ensemble                           & l+j+k         & 35      & 25     & 94,00\%   & 27,2    & 20,4   & 78,90\%   &                    \\
\textbf{n} &
  \textbf{Ensemble} &
  \textbf{j+k} &
  \textbf{36,6} &
  \textbf{27,4} &
  \textbf{94,40\%} &
  25 &
  19,5 &
  78,90\% &
  \textbf{NLE\_fa\_adhoc\_rr} \\
\textbf{o} &
  \textbf{Ensemble} &
  \textbf{a+b+c+d+g+h+i+j+k} &
  \textbf{34,9} &
  \textbf{25,3} &
  \textbf{93,50\%} &
  \textbf{28,1} &
  \textbf{20,7} &
  \textbf{76,70\%} &
  \textbf{NLE\_ru\_adhoc\_rr} \\ \bottomrule
\end{tabular}
}
\end{table}

\section{TREC NeuCLIR 2022 - initial analysis}
TREC Results are made available in Table~\ref{tab:trec}. We drew some initial conclusions and are still analyzing the results:

\begin{itemize}
\item In the end the dev set we used was not that good. For example on the first stage runs, the best run (mono vs AdHoc) was the exact opposite of the devset.
\item The Farsi monolingual reranked run was worse than the not reranked one, kinda the opposite of the dev set.
\item The gap between the baselines and the first stage runs was increased in the TREC set, further showing the advantage of ensembling first-stage rankers.
\item Of the reranked runs, the Adhoc was always better, achieving good results compared to the rest of the runs.
\item The mAP achieved in the Russian TREC set and the Recall@1k for both sets seem to confirm that our models are good at selecting good candidates, but are not as good at pushing them to the top. This is similar to what we observed in TREC DL 21 and 22.
\item Looking at the queries, there is a very large gap on how they are structured and how MSMARCO ones are. This probably impacted most runs of the competition, but it is especially critical on ours, where all runs are based on MSMARCO. The inclusion of training on HC4 is a probable step for next year.
\item Analyzing the hardest queries (c.f. Appendix), it seems that one problem was looking into the ``wrong'' database, by looking at questions for one nation/nationality/language on another. This makes us think of how a full-blown CLIR could happen (instead of English to one language, English to many languages).
\item Given the smaller gap in the effectiveness of mono vs adhoc in Farsi compared to Russian it is not unreasonable to imagine that it comes from the fact that machine translation from English to Farsi are worse than the English to Russian. Studying this is left as future work.

\end{itemize}

\begin{table}[ht]
\centering
\label{tab:trec}
\caption{Results for our group submissions. Bolded values are the best value for the subgroup}
\adjustbox{max width=\textwidth}{%
\begin{tabular}{c|cc|ccc|cc|ccc}
\toprule
\multirow{3}{*}{Name} & \multicolumn{5}{c|}{Farsi}                                    & \multicolumn{5}{c}{Russian}                            \\
                      & \multicolumn{2}{c|}{HC4}       & \multicolumn{3}{c|}{TREC}     & \multicolumn{2}{c|}{HC4} & \multicolumn{3}{c}{TREC}     \\
                      & NDCG@20       & mAP@1k        & NDCG@20 & mAP@1k & Recall@1k & NDCG@20     & mAP@1k    & NDCG@20 & mAP@1k & Recall@1k \\ \midrule
\multicolumn{11}{c}{Baselines}                                                                                                                \\ \midrule
splade\_\{\}\_mt &
  29.4 &
  20.5 &
  44.4 &
  \textbf{31.4} &
  \textbf{83.50\%} &
  21.6 &
  \textbf{16.8} &
  40.8 &
  29.3 &
  75.90\% \\
splade\_\{\}\_dt &
  27.4 &
  18.3 &
  \textbf{45.5} &
  28.7 &
  83.35\% &
  \textbf{23.6} &
  16 &
  \textbf{46.7} &
  \textbf{35.0} &
  \textbf{82.26\%} \\
splade\_\{\}\_ht      & \textbf{31.6} & \textbf{22.1} & 41.5    & 29.1   & 82.45\%   & 22.5        & 16.5      & 43.2    & 30.2   & 77.76\%   \\ \midrule
\multicolumn{11}{c}{First Stage Runs}                                                                                                         \\ \midrule
NLE\_\{\}\_mono &
  \textbf{31.9} &
  \textbf{22.6} &
  52.3 &
  \textbf{40.0} &
  91.27\% &
  23.8 &
  17.4 &
  \textbf{50.8} &
  \textbf{39.1} &
  \textbf{91.67\%} \\
NLE\_\{\}\_adhoc &
  29.3 &
  20.4 &
  \textbf{52.5} &
  39.3 &
  \textbf{91.97\%} &
  \textbf{26} &
  \textbf{20.4} &
  49.0 &
  38.4 &
  87.69\% \\ \midrule
\multicolumn{11}{c}{Reranker runs}                                                                                                            \\ \midrule
NLE\_\{\}\_mono\_rr   & 35.6          & 26.1          & 47.4    & 34.1   & 91.27\%   & 25.8        & 17.8      & 55.0    & 42.9   & 88.49\%   \\
NLE\_\{\}\_adhoc\_rr &
  \textbf{36.6} &
  \textbf{27.4} &
  \textbf{55.1} &
  \textbf{40.7} &
  \textbf{92.64\%} &
  \textbf{28.1} &
  \textbf{20.7} &
  \textbf{56.5} &
  \textbf{47.3} &
  \textbf{89.78\%} \\ \bottomrule
\end{tabular}

}
\end{table}

We also include some query analysis in the appendix (quite long so we preferred to keep it outside the ``main'' analysis.

\section{Conclusion}
For the TREC NeuCLIR 22 competition, we submitted runs in an effort to compare monolingual and AdHoc strategies. From the results, it seems that there is an interest in focusing on monolingual retrievers, however, given the larger literature and experiments in English, the AdHoc strategy is cheaper and as good/better than monolingual. Also much to our dismay, document translation worked the best out of all tested strategies. Finally, we also noticed that there's a need for a better development set in order to continue the efforts in this direction.

\bibliographystyle{apalike}  
\bibliography{main}  

\appendix
\section{Initial query analysis (nDCG values come from an older version of the qrels)}

\subsection{Topic 0}

\paragraph{Why we selected it:} Highest difference between median and max nDCG@20 in Russian (0.8983 vs 0.2732).

\paragraph{Title:} Iranian female athletes refugees

\paragraph{Description: } I am looking for stories about Iranian female athletes who seek asylum in other countries.

\paragraph{Narrative:} Find articles that identify female athletes who refused to return to Iran after competition and slam government by seeking asylum from other countries. Relevant documents must identify the athlete by name, and spell out the reason for seeking asylum and if they joined Refugee Team in Olympic competition. ``Official'' or unofficial reasons are equally acceptable, as long as the document gives some reason that the athlete refused to return after the competition,name of the countries that accepted their request, and Iran government reaction.

\paragraph{Why we think it is hard:} This query is directly meant for Iranian data, which probably makes it more sparse in Russian-based data. This is confirmed by the number of related documents found (12 in Russian and 99 in Farsi). 

\paragraph{How did our methods perform:} Looking at our methods it seemed that we found most of the related docs (11 out of 12) and effectively place them at the top (0.88 nDCG for our best model). 

\subsection{Topic 4}

\paragraph{Why we selected it:} Worst median mAP for Farsi (only 0.0024).

\paragraph{Title:} Corruption during construction of Vostochny Cosmodrome

\paragraph{Description: } I am looking for reports on officials convicted of corruption in the construction of the Russian Space launch Complex at Vostochny in the Far East.

\paragraph{Narrative:} Find articles on corruption and conviction of Russian officials and businessmen involved in the construction of the Russian Space launch Complex in Vostochny in the Far East. Relevant articles are to include information on Individuals, their duties, the crimes they are accused of, and plans by Roscosmos to alleviate the damage they did to the project. Issues not attributed to corruption are not relevant.

\paragraph{Why we think it is hard:} As the previous query, this deals with information mostly from a Russian point of view, thus making it normal to be very sparse in Farsi. Again this is confirmed by the number of related documents found (12 in Russian and 10 in Farsi). Another problem is on the entity (Vostochny Cosmodrome), it poses problems both for monolingual (tokenization may break the name) and for Adhoc (translation may remove the name).

\paragraph{How did our methods perform:} Our monolingual Farsi first stage achieved the best mAP, but all our other runs suffered, especially the reranked ones (probably due to the entity problem). 

\subsection{Topic 18}

\paragraph{Why we selected it:} Worst median nDCG@10 on Farsi (0 median, 0.74 max).

\paragraph{Title:} The situation of myopia rate among K12 students in China

\paragraph{Description: } I am interested in finding information about the situation of myopia among K12 students in China.

\paragraph{Narrative:} Find articles carry information about the situation of myopia rate among K12 students in China. Relevant documents should include information about the current situation and/or changes of the myopia rate, factors that affect the eyesight of K12 students, and the trend. Articles about the prevention and treatment of myopia, the research in the area of vision health, are not considered relevant.

\paragraph{Why we think it is hard:} As per the previous queries, this searches for Chinese information in the Farsi database, thus making judgments very sparse (only 5 positives). However, it is not as hard on the Russian database, pointing possibly to a language drift (K12 could have another name in Farsi).

\paragraph{How did our methods perform:} Our monolingual runs completely failed on this query (0.08 and 0.09 nDCG), but our Adhoc reranked got a very good result (0.6510). Looking at the baselines we confirm that only the document-translated runs could get a good result, pointing to the treatment of the Farsi language as the probable cause of trouble in this query.

\subsection{Topic 19}

\paragraph{Why we selected it:} No related documents found in Russian and not present in Farsi

\paragraph{Title:} Cases of formaldehyde related pollution in household products

\paragraph{Description: } I am interested in finding information about the situation of myopia among K12 students in China.

\paragraph{Narrative:} Find articles carry information about cases of formaldehyde related pollution in household products in China. Relevant articles should mention specific events that involve the pollution with information about the harms resulted in. Articles that discuss about the pollution of excessive formaldehyde, the harms and remedies, without mentioning any specific cases, are not considered relevant. Articles that discuss other types of pollutions found in household products are not considered relevant

\paragraph{Why we think it is hard:} This query misses the ``In China'' part when looking at just the title, which can cause problems with runs that only include the title. Moreover, as per the previous queries, this searches for Chinese information in the Russian database. 

\paragraph{How did our methods perform:} No related documents were found, however, nDCG is not 0 which we found very weird.

\subsection{Topic 20}

\paragraph{Why we selected it:} Worst median mAP in Russian (median 0.0006	and max 0.2778)

\paragraph{Title:} Are AIDS patients discriminated against in China

\paragraph{Description: } I am interested in finding out if AIDS patients are discriminated against in China.

\paragraph{Narrative:} Find articles that carry information about whether AIDS patients in China face any discrimination in the country. Relevant articles would be those that provide information about the situation in China from the following perspectives: discussions about discriminations that are found to exist, cases of discriminations and efforts to eliminate discriminations. Articles that discuss about AIDS and AIDS patients from the perspectives of medicine and public health including treatment are not relevant. Articles about discriminations that AIDS patients face in other countries are not relevant.

\paragraph{Why we think it is hard:}  As per the previous queries, this searches for Chinese information in the Russian database and it is not even evaluated in Farsi. Also as before, tokenization may cause problems with AIDS (Especially with case insensitive, with AIDS the disease vs aids the word).

\paragraph{How did our methods perform:} Our methods suffered on this task, the monolingual run had the best effectiveness out of the methods, but far from the best overall (0.04), but the reranker made it lose. The reranked Adhoc is slightly better. Looking at the baseline runs, it seems that the problem came from the machine translation of the query, as both DT and HT had decent mAP (0.04)

\subsection{Topic 24}

\paragraph{Why we selected it:} Highest difference between median and max on both nDCG@20 and mAP in Farsi (Some methods reached mAP=1 and nDCG=1, but median is around 0.1 for both)

\paragraph{Title:} Wind energy in Russia

\paragraph{Description: } I am looking for articles on the growth of wind energy in Russia.

\paragraph{Narrative:} Find articles on the development and growth of wind energy in Russia. Include information on joint ventures with foreign firms for the production of wind turbines and related equipment. Include information on the planned outputs of windmill parks. Information on wind energy in other countries not involving cooperation with Russian entities is not required. Neither is information on other renewable or non-renewable energy resources in Russia.

\paragraph{Why we think it is hard:}  As per the previous queries, this searches for Russian information in the Farsi database. Also, methods that use the title only would miss the need of ``growth''.

\paragraph{How did our methods perform:} 

\subsection{Topic 26}

\paragraph{Why we selected it:} Worst best recall on Russian (81.58\%)

\paragraph{Title:} Ukrainian Presidential Candidate Zelenskiy

\paragraph{Description: } I am looking for articles that reflect Russia's attitude towards Ukrainian Presidential Candidate Volodymyr Zelenskiy

\paragraph{Narrative:} Find articles that reveal the Russian perspective of then presidential candidate Volodymyr Zelenskiy for the 2019 election. Include articles that highlight what Ukrainian officials say that Russian media circulates about Zelenskiy as that reveals what Russians want readers to be aware of. Russian perspective as to his ability to win, who might be supporting him, how seriously they should be taking him, and so forth. Did not include articles that discussed subsequent elections, local or parliamentary elections, or any actions Zelenskiy took once he became president. Did not include Zelenskiys recommendations on how elections in Donbass should be run. Articles purely about the polls and statistics were not included.

\paragraph{Why we think it is hard:} There's a problem with transliteration for translations, where term-based methods may miss documents due to it having written Zelenski or Zelensky or Zelenskyy (found all these variants after a quick google search). 

\paragraph{How did our methods perform:} Our final methods got decent results (78.95\% the best), but not because of SPLADE, which had very bad recall: DT: 21.05\% , MT:15.79\%, HT: 7.89\%.

\subsection{Topic 52}

\paragraph{Why we selected it:} Worst median nDCG@20 on Russian (0.1095, with a max of 0.6723) 

\paragraph{Title:} Tourism in Beijing under the Covid-19 pandemic

\paragraph{Description: } I am interested in finding information about the tourism industry in Beijing amid the Covid-19 pandemic.

\paragraph{Narrative:} Find stories about the situation of the tourism industry in Beijing under the Covid-19 pandemic. Relevant articles are those that would cover any aspect of the industry in the city, like the situation of museums and other attractions, restaurants, tour operators, etc. during the time when the pandemic remained a concern. Stories about outbound travel from Beijing would not be considered relevant. Only the actual situations would be of interest for the information search. Articles of analysis of the industry and projections, unless they do mention the actual situation, would not be considered relevant

\paragraph{Why we think it is hard:}  Again, the title misses a keyword (industry) and asks for details of China in a Russian base, with the Farsi base also not as effective (0.2321, with a max of 0.5379). 

\paragraph{How did our methods perform:} Our best method was almost as effective as the best (0.59 vs 0.67), but looking into our first stage rankers, we have a very large drop on the document back-translation (0.2551) compared to the query translations (0.5334)

\subsection{Topic 67}

\paragraph{Why we selected it:} Highest difference between median and max mAP on Russian (max mAP=1, median mAP=0.0606)

\paragraph{Title:} Reasons people become virtual streamers

\paragraph{Description: } I'm looking for reasons why people choose to stream online virtually behind an avatar.

\paragraph{Narrative:} Find articles explaining why people choose to stream online virtually behind a 2D or 3D avatar. Relevant articles shoud provide attribute or positive aspects of streaming virtually that are superior to traditional live-person broadcast.

\paragraph{Why we think it is hard:} This is the first topic from the set we analyzed that is not dependent on a language (or focused on a local subject). The lack of the Vtuber (or V-tuber) keyword could be something problematic here, but actually, the main problem is that there's only one positive. 

\paragraph{How did our methods perform:} All methods found the only positive, but only the SPLADE DT had it at the first position, with the other methods putting it at best at 5th.

\subsection{Topic 72}

\paragraph{Why we selected it:} Not evaluated in Farsi

\paragraph{Title:} Playing games could effectively prevent dementia

\paragraph{Description: } I'm looking for articles on whether playing games could prevent or delay the onset of dementia.

\paragraph{Narrative:} Find articles that mention playing games could prevent or delay the onset of dementia. Table top games and video games are equally relevant. Relevant articles should provide examples of games being played that have shown evidance of cognitive maintanance. Relevant articles should also mention what type of cognitive activity or region of brain these games positively stimulate.

\paragraph{Why we think it is hard:}  Like the previous one, this one does not necessarily focus on a particular language, but it could be that no results exist on the database, as it seems to be the case for Farsi. In Russian, there were some results.

\subsection{Topic 118}

\paragraph{Why we selected it:} Worst best nDCG@20 on Russian (0.4967, with a median of 0.2607) and also the worst best Recall@1k on Farsi (85.02\%).

\paragraph{Title:} Arbitrary detention of foreigners and Iranian dual nationals by Iran

\paragraph{Description: } I'm looking for articles reporting on the arbitrary detention of foreigners and Iranian dual nationals by Iran.

\paragraph{Narrative:} Find articles that show whether the arbitrary detention of foreign nationals and dual nationals by Iran is for political or economic gain or whether they are allegedly spies for Iran. What is the reaction of governments and international institutions? Is this a government hostage-taking.

\paragraph{Why we think it is hard:} Again, we have the problem of looking for pieces of information of ``one'' base in another. However, it also ``caused problems'' on Farsi, but for the opposite reason: there were too many positives (287). 

\paragraph{How did our methods perform:} In Farsi we are able to achieve the best possible Recall on the Adhoc but slightly less on the monolingual task. While in Russian we are almost able to match the best nDCG@20 (0.4955 vs 0.4967) and have similar effects on both Adhoc and monolingual.

\end{document}